\documentclass[12pt]{article}
\usepackage[dvips]{color}
\usepackage{epsfig}
\usepackage{graphicx}
\usepackage{amsmath}

\begin{document}
\begin{center}
\Large{\bf The equivalence between Finsler and non-commutative geometries by massive gravity black hole}\\
\small \vspace{1cm} {\bf J. Sadeghi $^{\dag}$
\footnote{pouriya@ipm.ir}}, \quad{\bf Z. Nekouee $^{\ddag}$
\footnote{z.nekouee@stu.umz.ac.ir}} \quad and {\bf A. Behzadi $^{\ddag}$
\footnote{behzadi@umz.ac.ir}}\\
\vspace{0.5cm}$^{\dag}${\it Department of Physics, Faculty of Basic Sciences,\\
University of Mazandaran, P. O. Box 47416-95447, Babolsar, Iran}\\
\vspace{0.5cm}$^{\ddag}${\it Department of Mathematic, Faculty of Mathematic,\\
University of Mazandaran, P. O. Box 47416-95447, Babolsar, Iran}\\
\end{center}\vspace{1.2cm}
\date{\today}

\begin{abstract}
In the present work, we wanted to find the possible way in order to make the equivalence between
non-commutative and Finsler geometries as two useful mathematical tools. Based on this purpose, we were concerned to search this possibility by investigating the massive gravity black holes. Firstly the Lagrangian of the system is introduced and then it is rewritten in the non-commutative regime by definition of the new variables. On the other hand, we focus on the Finsler geometry in order to find a Finslerian function which is equivalent to the mentioned non-commutative Lagrangian under special conditions. Also, the effective potential of the system was calculated as a part of the corresponding conditions.
\end{abstract}
\section{Introduction}
One of the most interesting parts of mathematical physics is non-commutative ($NC$) geometry which is applied in many branches of physics [1,2]. In principle, the connection between this geometry and field theory plays an important role for explantation of Hall effects. From $NC$ geometry point of view, in order to make quantum gravity ($QG$) as a gravitational theory in very early universe, we can engage the various versions of the $NC$ geometry [3]. For example, in order to explain gravity in the context of quantum mechanics, string theory employs the non-commutativity for the special coordinates instead of either spatial or time [4-6]. We note that here if we consider both coordinates, it may be lead to some difficulties for unitarian and causality. In contrast, there are some $QG$ models which are established on the classical gravity with quantum corrections. As an important model, loop quantum gravity ($LQG$) uses $NC$ geometry in the discrete spaces against the canonical quantum gravity ($CQG$) model which has continues space in commutative background. Another $QG$ approach is associated with the application of $NC$ geometry for $CQG$ with the deformed space-time as the main attitude in the present paper. This kind of non-commutativity is applied to our recent cosmological papers [7-9].\\
On the other hand, we want to present some explanations about Finsler as a second target geometry. Finsler geometry was first applied to gravitational theory, this led to correct the results predicted by general relativity [10-18]. As we know general relativity ($GR$) as gravitational theory in classical scales is founded on the Riemannian geometry and also its corresponding cosmological model is endorsed by remarkable observational evidences. Some ambiguities in cosmological constant and other problem in cosmology lead us to apply the generalized form of the Riemannian geometry as Finsler geometry to overcome our obstacles. So, here we focus on Finsler geometry. It can open the new window to the research.\\
As we know, modern physics is established upon the local Lorentz symmetry and CPT invariance. The theoretical
investigation and experimental examination of Lorentz
symmetry have made considerable progress [19]. On the other hand,
there are many attempts to investigate the possible
Lorentz violation from theoretical aspect [20]. Coleman and
Glashow consider the case of space-time translations along
with exact rotational symmetry in the rest frame of the
cosmic background radiation. But this situation leads us to have a small departures
from boost invariance in the corresponding frame. Here, in order to have suitable departure of Lorentz invariance they developed perturbation and such invariance is parameterized in terms of fixed time-like four-vector [21]. Colladay and Kostelecky
proposed the model incorporating Lorentz and CPT violation
extension of the standard model by introducing
into the Lagrangian [22]. It is also argued that the large boosts
naturally uncover the structure of space-time at arbitrary
small scales. In that case, we need to modify the corresponding theory with an additional fundamental length as a  Planck scale.
Modification of special relativity
is known as doubly
special relativity (DSR)[23]. The realization of DSR can be non-linear realization of Poincar\'{e} group. Such realization covers the structure of space-times which is complectly related to $NC$ space-time [24]. The main feature of these
realizations of DSR involves the deformed dispersion relation
which can also lead to Finsler type of space-time
geometry [25]. So, we take advantage of above relation and investigate the equivalency between $NC$ and Finsler geometry. For this reason, in the following section, firstly we are going to explain the Finsler geometry.\\
Let's consider the features of this geometry. From mathematical approach length of a vector $\|x\|$ in a manifold $M$ endowed by a metric $g_{ij}(x)$ in Riemannian geometry is given by the quadratic form,
\begin{equation}
\|x^{2}\|=g_{ij}(x)x^{i}x^{j}
\end{equation}
where $i$ and $j$ denote to spatial components. Moreover, Finsler geometry can present a more general method to determine
the norms of the vectors. Because of this reason, Finsler geometry is a generalization of Riemaniann geometry. In this geometry, the length of the vectors can be defined by a general method which is not confined by Riemaniann definition of length.

According to the mathematical definition of this geometry, we can introduce an $N$ dimensional manifold $M$ which is supplied by a positive scalar function $F(x,y):TM\rightarrow[0,\infty)$ where $x=x^{i}=(x^{0}, ...., x^{n})\in M$, $y=y^{i}=(y^{0}, ...., y^{n})\in T_{x}M$ and $TM$ is the tangent bundle. Suppose that $x_{i}$ is the local coordinate on $M$. Then the local coordinate on $TM$ is shown by $(x_{i}, y_{i})$ where $y_{i}$ is the fiber coordinate. With respect to local coordinate system induced on $TM_{0}$, the natural local frame fields on $TM_{0}=TM-\{0\}$ are given by
$\dot{\partial_{i}}=\frac{\partial}{\partial y_{i}}$ and $\partial_{i}=\frac{\partial}{\partial x_{i}}$. The vertical distribution $VTM_{0}$ is locally spanned by $\dot{\partial_{i}}$, $i = 1,..., n$. Considering the fundamental function $F$, then the horizontal distribution $HTM_{0}$ as a complementary distribution of $VTM_{0}$ can be naturally defined. Finsler metric function have to satisfy three properties: The first one is regularity which $F(x,y)$ is differentiable on the slit tangent bundle $TM_{0}$. The second one is positive homogeneity, e.g. $F(x,y)$ is homogenous function of degree one in $y$ as $F(x,\lambda y)=\lambda F(x,y)$ for any number of positive $\lambda$. The third one is strong convexity, it can be found that the quadratic form
$\frac{\partial^{2}F(x,y)}{\partial y^{i}\partial y^{j}}y^{i}y^{j}$ is assumed to be positive definite for all variables [26-28]. The study of physical Lagrangians in Finsler regime is very interesting and practical in our analysis. We note that our set up will be restricted to a class of gauge theory Lagrangians which are equipped with weak regularity and positive homogeneity conditions of Finsler geometry.\\

So far, we have introduced two useful geometries $NC$ and Finsler which can play a great role in physical situations. The main aim of this paper is associated to find the possible connection between these geometries in physical systems. Because of this, we attempt to supply our purpose by focusing on massive gravity black holes. $GR$ as classical gravitational theory is a relativistic theory where the graviton
is massless. One can ask that whether we can fabricate a self-consistent gravity theory
if graviton is massive?. It is obviously, it's  hard to answer because of theoretical difficulties. In Ref. [29-31], classical theories of nonlinear massive gravity has been proposed in the absence of ghost field [32,33]. Also, a nontrivial black hole solution with a Ricci
at horizon in $4D$ massive gravity is reported in Ref. [34] with a negative cosmological constant [35]. New methods for phase transition of charged black holes in massive gravity are introduced in Ref. [36]. Some holographic consequence of the effect of graviton mass in massive gravity has been investigated in Ref. [37-39]. Also, the fluid/gravity correspondence in the context of massive gravity is investigated [40]. Likewise, in Ref. [41] the black hole solutions of massive gravity are investigated in a good framework. The thermodynamics of massive gravity black holes has been considered in Ref. [42]. In addition, in Ref. [43,44], some thermodynamical features of AdS black holes have been presented in the context of massive gravity. In Ref. [45], covariant renormalizable modified and massive gravity theories are studied on (Non) commutative tangent Lorentz bundles. In the present work, we focus on the Lagrangian of the massive gravity black hole as a target Lagrangian in order to find the possible link between $NC$ and Finsler geometries. This gives us motivation to arrange the paper as following strategy. In section $2$, we present a small review for massive gravity black holes and it will be followed by the application of $NC$ geometry for the mentioned black holes in section $3$. In section $4$, we present some calculations in  frame of Finsler geometry. In conclusion, by comparing the results in the previous sections, we will show that one can find a kind of equivalence between $NC$ and Finsler geometries under some conditions. Also, we find an expression for the effective potential of the physical system as a part of our requirements.
\section{The massive black hole solution}
First of all, we consider the action for a four-dimensional Einstein-Maxwell theory in the framework of massive gravity [46-48]. In that case, we need to present its corresponding black hole solution. By using the conventions of Refs. [49-54], one can write the massive Einstein-Maxwell action consisting of the Ricci scalar, electromagnetic field, cosmological constant term, graviton mass terms and a surface term [55-57], which is given by,
\begin{equation}
S=\int d^{4}x\sqrt{-g}[\frac{1}{2\kappa^{2}}(R-2\Lambda)
+\frac{m^{2}}{\kappa^{2}}(\alpha_{1}u_{1}+\alpha_{2}u_{2})-\frac{1}{16\pi}F^{2}]+\frac{1}{\kappa^{2}}\int d^{3}x\sqrt{-\gamma}K,
\end{equation}
where
\begin{equation}
u_{1}=tr\mathcal{K},
\end{equation}
\begin{equation}
u_{2}=(tr\mathcal{K})^{2}-tr(\mathcal{K}^{2}),
\end{equation}
$\alpha_{1},\alpha_{2}\leq 0$, $\kappa^{2}=8\pi G$, $\mathcal{{K^{\mu}}_{\nu}}:=\sqrt{g^{\mu\alpha}f_{\alpha\nu}}$. The reference metric without dynamic behavior is chosen as $f_{\mu\nu}=diag(0,0,1,\sin^{2}\theta$) and $K$ is the trace of the extrinsic curvature. In order to have some Lagrangian for the massive gravity system, we need black hole solution of action (2) with gauge field $A_{\mu}$  which are given by,
\begin{equation}
ds^{2}=-f(r)dt^{2}+\frac{dr^{2}}{f(r)}+r^{2}d\theta^{2}+r^{2}\sin^{2}\theta d\varphi^{2},
\end{equation}
\begin{equation}
A_{\mu}=(A_{t},0,0,0).
\end{equation}
By variation of the action (2) with respect to the metric and using the Eq. (5), we can find field as,
\begin{eqnarray}
R_{\mu\nu}-\frac{1}{2}Rg_{\mu\nu}-\frac{3}{l^{2}}g_{\mu\nu}+
m^{2}\alpha_{1}(\mathcal{K}_{\mu\nu}-tr\mathcal{K}g_{\mu\nu})
+m^{2}\alpha_{2}[2(tr\mathcal{K})\mathcal{K}_{\mu\nu}-2{\mathcal{K}_{\mu}}^{\alpha}\mathcal{K}_{\alpha\nu}&\!]&\!\nonumber\\
-m^{2}\alpha_{2}g_{\mu\nu}[(tr\mathcal{K})^{2}-tr(\mathcal{K}^{2})]=
2G(F_{\mu\alpha}{F_{\nu}}^{\alpha}-\frac{1}{4}g_{\mu\nu}F^{2}),
\end{eqnarray}
\begin{equation}
\nabla_{\mu}F^{\mu\nu}=0,
\end{equation}
where $f(r)$ and $A_{t}$ can be defined as following terms,
\begin{equation}
f(r)=1-\frac{2GM}{r}+\frac{GQ^{2}}{r^{2}}+\frac{r^{2}}{l^{2}}+m^{2}\alpha_{1}r+2m^{2}\alpha_{2},
\end{equation}
\begin{equation}
A_{t}=-\frac{Q}{r}.
\end{equation}
where $M$ and $Q$ are mass and total charge of black hole.\\
In order to obtain our main Lagrangian, we can employ  general form of Lagrangian as,
\begin{equation}
\mathcal{L}=\frac{1}{2}(g_{\mu\nu}\frac{dx^{\mu}}{d\tau}\frac{dx^{\nu}}{d\tau})+eA_{\mu}\frac{d x^\mu}{dt}.
\end{equation}
where  parameter $\tau$ is the proper time of the charged test particle. The components of this metric can be found as
\begin{equation}
A_{t}=-\frac{Q}{r},
\end{equation}
and
\begin{equation}
\begin{array}{ccc}
g_{tt}=-f(r), & g_{rr}=\frac{1}{f(r)}, & g_{\varphi\varphi}=r^{2}.
\end{array}
\end{equation}
Moreover, according to our purpose to attain the connection between $NC$ and Finsler geometries, we restrict our analysis to the case of non-charged test particle ($(Q=0)$) in the equatorial plane with $\theta=\frac{\pi}{2}$, $d\theta=0$ and $\sin\theta=1$. So, the form of  metric can be reduced as,
\begin{equation}
ds^{2}=-f(r)dt^{2}+\frac{dr^{2}}{f(r)}+r^{2}d\varphi^{2},
\end{equation}
also the Lagrangian of black hole takes the following form,
\begin{equation}
\mathcal{L}=\frac{1}{2}(g_{\mu\nu}\frac{dx^{\mu}}{d\tau}\frac{dx^{\nu}}{d\tau}).
\end{equation}
In order to obtain the geodesic equations, we define the conjugate momentum $P_{\mu}$ to the coordinate $x^{\mu}$ as,
\begin{equation}
P_{\mu}\equiv\frac{\partial\mathcal{L}}{\partial \dot{x}^{\mu}}=g_{\mu\nu}\dot{x}^{\nu},
\end{equation}
where
\begin{equation}
\dot{x}^{\mu}\equiv\frac{dx^{\mu}}{d\tau}=u^{\mu}.
\end{equation}
Now, we use the Eqs. (14) and (15) in case of $Q=0$ and obtain the corresponding Lagrangian for the massive gravity as,
\begin{equation}
\mathcal{L}=\frac{1}{2}(-f(r)\dot{t}^{2}+\frac{1}{f(r)}\dot{r}^{2}+r^{2}\dot{\varphi}^{2}).
\end{equation}
The Euler-Lagrange equations can be written In terms of conjugate momenta.
\begin{equation}
 \frac{d}{d\tau}P_{\mu}=\frac{\partial\mathcal{L}}{\partial x^{\mu}}.
 \label{eqn19}
\end{equation}
The above black hole has two Killing vectors: a time-like Killing vector $k^{\mu}=(1,0,0,0)$
and space-like Killing vector $m^{\mu}=(0,0,0,1)$. Due to the space-time symmetry, two conserved quantities are represented by two Killing vectors $\partial_t$ and $\partial_\phi$. Here,
we note that the massive gravity follows time-like geodesic along the two above quantities. These constants are given by $E=-\partial_t. U$ and $L=\partial_\phi. U$
where $U$, $E$ and $L$ correspond to the four-velocity of the massive gravity, energy and angular
momentum test particle, respectively. In this study, as we pointed out, we focus on time-like
geodesics in the equatorial plane $(\theta =\frac{\pi}{2})$ implying
$U^\theta=0 $.  The radial motion of the
massive particle in the metric is obtained by solving the equation
$U.U=-1.$
Hence, the canonical momenta $P_t$ and $P_{\phi}$ are conserved and these constants are labeled as energy
per unit mass $E$ and angular moment per unit mass $L$. In that case the conserved quantities are given by,
\begin{equation}
E\equiv -k_{\mu}u^{\mu}=-g_{t\mu}u^{\mu}=-P_{t},
\label{eqn20}
\end{equation}
\begin{equation}
L\equiv m_{\mu}u^{\mu}=-g_{\varphi\mu}u^{\mu}=P_{\varphi},
\label{eqn21}
\end{equation}
where $E$ is the energy at infinity and $L$ is the angular momentum. In order to obtain $\dot{t}$ and $\dot{\varphi}$. We have to consider Eqs. (16) and (19), so we obtain the following equation,
\begin{equation}
 \dot{t}=\frac{E}{f(r)}=u^{t},
\end{equation}
\begin{equation}
\dot{\varphi}=\frac{L}{r^{2}}=u^{\varphi}.
\end{equation}
To find the effective potential, we employ the following equation,
\begin{equation}
g_{\mu\nu}u^{\mu}u^{\nu}=\kappa,
\end{equation}
where $\kappa=-1$ for time-like geodesics. So, by combination of Eqs. (22) to (24) and $\dot{r}^{2}+V_{eff}=0$, the $V_{eff}$ can be expressed as
\begin{equation}
V_{eff}=f(r)(\frac{L^{2}}{r^{2}}+1)-E^{2}.
\end{equation}
In the next section, we want to review $NC$ definition and also obtain some new change of variables for the deformation theory. Here, we are going to explain some physical motivations for the relation between Finsler and $NC$ geometry.
As we know, when we deformed the gravity theory one could arrive at some modified dispersion relations. Also, there is already a well-established common point between the $DSR$-relativistic theories framework and Finsler geometry. This relation leads us to modified the corresponding theory from $NC$ approach. The obtained modified massive gravity theory is the possibility of allowing for modified dispersion relations. So, the connection between the $DSR$ and Finsler geometry as the above mentioned leads us to arrange the relation between Finsler and $NC$ geometry. Now we go back to reviewing the $NC$ approach and obtain the corresponding Lagrangian.\\
 \section{The massive gravity Lagrangian in non-commutative geometry}
In this section, we write the Lagrangian of massive gravity in Eq. (18) in the framework of $NC$ geometry. In that case, the canonical relations help us to obtain the corresponding Hamiltonian. In order to apply $NC$ geometry to the pointed Hamiltonian, we need to rewrite the Hamiltonian in terms of new variable. In that case, the new Hamiltonian for the Lagrangian of massive gravity will be form of simple harmonic oscillator. The mentioned strategy can be applied as following calculations. First, we try to choose following variables,
\begin{equation}
\begin{array}{ccc}
x_{1}=\sqrt{f(r)}\cosh t, & x_{2}=r\sinh\varphi, & x_{3}+y_{3}=\sqrt{2}r+\sqrt{f(r)},\\
y_{1}=\sqrt{f(r)}\sinh t, & y_{2}=r\cosh\varphi, & x_{3}-y_{3}=\sqrt{2}r-\sqrt{f(r)},\\
\end{array}
\end{equation}
and
\begin{equation}
\begin{array}{cc}
x_{4}=\frac{1}{\sqrt{f(r)}}\sinh r, & x_{5}+y_{5}=\frac{1}{\sqrt{f(r)}}+r,\\
y_{4}=\frac{1}{\sqrt{f(r)}}\cosh r, & x_{5}-y_{5}=\frac{1}{\sqrt{f(r)}}-r.
\end{array}
\end{equation}
The Hamiltonian is,
\begin{equation}
H=\frac{1}{2}\sum\limits_{i=1}^5 ((P_{x_{i}}^{2}-P_{y_{i}}^{2})+\omega_{i}^{2}(x_{i}^{2}-y_{i}^{2})),
\end{equation}
where
\begin{equation}
\begin{array}{c}
P_{x_{i}}=\frac{\partial\mathcal{L}}{\partial\dot{x_{i}}}=\dot{x_{i}},\\
P_{y_{i}}=\frac{\partial\mathcal{L}}{\partial\dot{y_{i}}}=-\dot{y_{i}},
\end{array}
\end{equation}
and
\begin{eqnarray}
\omega_{i}^{2}=-1.
\end{eqnarray}
It is obvious that the Hamiltonian Eq. (28) has an oscillator form that
is useful for gravitational theories. Before applying $NC$ geometry to the Hamiltonian (28), one may need to review some features such geometry.
As we know in commutative case we have usual Poisson brackets which are given by,
\begin{equation}
\{x_{i},x_{j}\}=0,\quad\{P_{x_{i}},P_{x_{j}}\}=0,\quad\{x_{i},P_{x_{j}}\}=\delta_{ij},
\end{equation}
where $x_{i}(i=1, 2)$ and $P_{x_{i}}(i=1, 2)$.  To compare $NC$ and commutative phase space
we need to explain some approaches of $NC$. In such an
approach, quantum effects can be dissolved by the Moyal brackets
$\{f,g\}_{\alpha}=f\star_{\alpha}g-g\star_{\alpha}f$ which is based
on the Moyal product as,
\begin{equation}
(f\star_{\alpha}g)(x)=exp[\frac{1}{2}\alpha^{ab}\partial_{a}^{(1)}\partial_{b}^{(2)}]f(x_{1})g(x_{2})|_{x_{1}=x_{2}=x}.
\end{equation}
After corresponding calculations, we find the algebra of variables as,
\begin{equation}
\{x_{i},x_{j}\}_{\alpha}=\theta_{ij},\quad\{x_{i},P_{j}\}_{\alpha}=\delta_{ij}+\sigma_{ij},\quad\{P_{i},P_{j}\}=\beta_{ij}.
\end{equation}
Transformations on the classical phase space variables are expressed
as,
\begin{equation}
\hat{x_{i}}=x_{i}+\frac{\theta}{2}P_{y_{i}},\quad \hat{y_{i}}=y_{i}-\frac{\theta}{2}P_{x_{i}},\quad \hat{P_{x_{i}}}=P_{x_{i}}-\frac{\beta}{2}y_{i},
\quad \hat{P_{y_{i}}}=P_{y_{i}}+\frac{\beta}{2}x_{i},\\
\end{equation}
The deformed algebra for new variables are given by,
\begin{equation}
\{\hat{y},\hat{x}\}=\theta,\quad\{\hat{x},\hat{P_{x}}\}=\{\hat{y},\hat{P_{y}}\}=1+\sigma,\quad\{\hat{P_{y}},\hat{P_{x}}\}=\beta,
\end{equation}
where $\sigma=\frac{\beta\theta}{2}$. In order to construct the
deformed Hamiltonian of massive gravity, we borrow the form of Hamiltonian
from Eq. (28) with new variables in Eq. (34). Hence, the form of
Hamiltonian in the deformed analysis is found as,
\begin{equation}
\hat{H}=\frac{1}{2}\sum_{i=1}^{5}((P_{x_{i}}^{2}-P_{y_{i}}^{2})-\gamma_{i}^{2}(y_{i}P_{x_{i}}+x_{i}P_{y_{i}})+
{\tilde{\omega_{i}}^{2}}(x_{i}^{2}-y_{i}^{2})),
\end{equation}
where
\begin{equation}
{\tilde{\omega_{i}}}^{2}=\frac{{\omega_{i}}^{2}-\frac{\beta^{2}}{4}}{1-{\omega_{i}}^{2}
\frac{\theta^{2}}{4}},\quad
{\gamma_{i}}^{2}=\frac{\beta-{\omega_{i}}^{2}\theta}{1-{\omega_{i}}^{2}\frac{\theta^{2}}{4}}.
\end{equation}
The above deformed Hamiltonian lead us to have a new Lagrangian which is deformed form of original Lagrangian of massive gravity.
Now, for the corresponding
model we arrange the deformed Lagrangian as,
\begin{equation}
\hat{\mathcal{L}}=\frac{1}{2}\sum_{i=1}^{5}((P_{x_{i}}^{2}-P_{y_{i}}^{2})-{\hat{\gamma_{i}}^{2}}(y_{i}P_{x_{i}}+x_{i}P_{y_{i}})-
{\hat{\omega_{i}}^{2}}(x_{i}^{2}-y_{i}^{2})),
\end{equation}
where
\begin{equation}
{\hat{\omega_{i}}}^{2}=\frac{{\omega_{i}}^{2}+\frac{\beta^{2}}{4}}{1+{\omega_{i}}^{2}
\frac{\theta^{2}}{4}},\quad
{\hat{\gamma_{i}}}^{2}=\frac{\beta+{\omega_{i}}^{2}\theta}{1+{\omega_{i}}^{2}\frac{\theta^{2}}{4}}.
\end{equation}
We use the Eq. (29), then the deformed Lagrangian can be written as following,
\begin{eqnarray}
\hat{\mathcal{L}}=\frac{1}{2}(-f(r)\dot{t}^{2}+\frac{1}{f(r)}\dot{r}^{2}+r^{2}\dot{\varphi}^{2}
-\frac{(\beta-\theta)}{1-\frac{\theta^{2}}{4}}(-f(r)\dot{t}+r^{2}\dot{\varphi}&\!-&\!\nonumber\\
(\sqrt{2f(r)}+\frac{1}{f(r)}-\frac{1}{\sqrt{f(r)}}-\frac{\sqrt{2}}{2}\frac{\acute{f}(r)r}{\sqrt{f(r)}}
-\frac{\acute{f}(r)r}{2f(r)\sqrt{f(r)}})\dot{r})),
\end{eqnarray}
where
\begin{equation}
  \acute{f}(r)=\frac{d}{dr}f(r).
\end{equation}
In the next section we have to explain Finsler geometry from physical and mathematical point of view.
\section{The massive gravity Lagrangian in Finsler geometry}
Before we define some aspects of such geometry one can find more physical motivations for the Finsler geometry. Lorentz invariance plays an important role in the standard model of physics. Also, the Lorentz invariance is investigated by experiments and theories. The theoretical approach of investigating the Lorentz invariance is studying the space-time symmetry. Such space-time symmetry leads us to introduce some counterparts of special relativity. The first one is $DSR$ [23]. In these counterparts, we need two invariant parameters as \textit{k} and \textit{c} which are Planckin  and speed of light parameters respectively. The last one is the de Sitter ($dS$)/anti-de Sitter ($AdS$) invariant special relativity ($dSSR$) [58,59]. Here we note that the $dSSR$ suggests  the principle of relativity should be generalized to constant curvature space-time with radius R in Riemannian geometry. Snyder proposed a quantized space-time model [60]. In this model, he defined space-time coordinates as transition generators of so(1,4) algebra, and become $NC$ geometry. There is one-to-one correspondence between Snyder's quantized space-time as a $DSR$ and the $dSSR$ models. Also, $VSR$ can be realized on $NC$ Moyal plane with light-like non-commutativity [61]. On the other hand, these counterparts of special relativity have connections with Finsler geometry [26], which is a natural
generalization of Riemannian geometry. The non-commutativity effects may be regarded as the deviation of Finsler space-time
from Riemann space-time. So all these connections give us some motivations to relate between Finsler and $NC$ geometry with the massive gravity background. In this section first we try to explain some aspects of Finsler geometry and find the corresponding Lagrangian.
Now we can concentrate our attention on Finsler geometry issues and also we take advantage of Lagrangian in massive gravity system. Firstly, we are going to review some properties in Finsler geometry. So, here we note that, a Finsler space $F^{n}=(M,F)$ is an $n$-dimensional manifold $M$
equipped with a Finsler metric $ F:TM\rightarrow{\rm I\!R}$,
$(x,y)\longmapsto F(x,y)$, $x\in M$, $y\in TM$, with following conditions. First condition is regularity
where  $F$ is a $C^{\infty}$ function on $TM\backslash\{0\},$
second condition is positive homogeneity (of degree 1) where $F(x,\lambda Y)=\lambda
F(x,y), \lambda\in{\rm I\!R} $
and third condition is  strong convexity where
$g_{ij}(x,y)=\frac{1}{2}\frac{\partial^{2}F^{2}}{\partial
y^{i}\partial y^{j}}(x,y)$ is positively defined.\\
Now we are going to introduce the general form of metric in Finsler geometry which is given by following,
\begin{equation}
 F(x,y)=\alpha a(x,y)+\eta b(x,y)+\gamma \frac{a^{2}(x,y)}{b(x,y)}+\Xi(x,y),
\end{equation}
 where $a(x,y)=\sqrt{a_{ij}(x)y^{i}y^{j}}$ is a Riemannian
metric, $a^{ij}$ is the
inverse matrix of $a_{ij}$  and $b(x,y)$ is a differential one-form on $M$ with
$\|b(x,y)\|_{a}:=\sqrt{a^{ij}b_{i}b_{j}}<1$. Here, we used  Einstein notation and such equation (45), defines a Finslerian metric. Also one can say that
$\Xi(x,y)$ is a Finsler fundamental function on $TM$  and $\alpha, \eta, \gamma\in\mathcal{F}(M)$. The first two terms of $F$  in above metric determine a Randers metric. Randers metrics are among the simplest Finsler metrics. These metrics were first introduced by physicist G. Randers from the standpoint of general relativity [62]. Later on, these metrics were applied to the theory of the electron microscope by R. S. Ingarden in [63], who first named them Randers metrics. Up to now, many Finslerian geometers have made great efforts in investigation on the geometric properties of Randers metrics [64,65]. The Finsler metric (Lagrangian) corresponding to massive gravity solution without charge will be as,
\begin{eqnarray}
F(x,y)=\alpha\sqrt{-f(r)\dot{t}^{2}+\frac{1}{f(r)}\dot{r}^{2}+r^{2}\dot{\varphi}^{2}}&\!+&\!\nonumber\\\eta
(-f(r)\dot{t}+r^{2}\dot{\varphi}+(\sqrt{2f(r)}+\frac{1}{f(r)}-\frac{1}{\sqrt{f(r)}}-\frac{\sqrt{2}}{2}
\frac{\acute{f}(r)r}{\sqrt{f(r)}}-\frac{\acute{f}(r)r}{2f(r)\sqrt{f(r)}})\dot{r}).
\end{eqnarray}
Now, we compare such metric to Eq. (42) one can suppose that $\Xi=\gamma=0$. The most important result here is the comparison of obtained results of Lagrangian between Finsler and $NC$ geometry. In order to compare such results in Eqs. (40) and (43) we have to apply following conditions to Eq. (43),
\begin{equation}
 \dot{r}^{2} \ll V_{eff},\qquad \alpha=\sqrt{\frac{l^{2}}{r^{2}}-\frac{E^{2}}{f(r)}}.
\end{equation}
Also, we note that the system with mentioned conditions has a stability as ${\omega_{\varphi}}^{2} \gg 1$.
According to the mentioned conditions $F(x,y)$ can be written by,
\begin{eqnarray}
F(x,y)\simeq(-f(r)\dot{t}^{2}+\frac{1}{2}\frac{1}{f(r)}\dot{r}^{2}+r^{2}\dot{\varphi}^{2})&\!+&\!\nonumber\\\eta
(-f(r)\dot{t}+r^{2}\dot{\varphi}+(\sqrt{2f(r)}+\frac{1}{f(r)}-\frac{1}{\sqrt{f(r)}}-\frac{\sqrt{2}}{2}
\frac{\acute{f}(r)r}{\sqrt{f(r)}}-\frac{\acute{f}(r)r}{2f(r)\sqrt{f(r)}})\dot{r}).
\end{eqnarray}
Here, we see some equivalencies between the results of two geometries in massive gravity background. By using the condition as $\eta=\frac{(\theta-\beta)}{1-\frac{\theta^{2}}{4}}$ for $\theta^{2}\simeq0$, the Lagrangian of massive gravity solution from  $NC$ and Finsler geometries will be same. But in case of $\theta=\beta$, we found that they are not any deformation to the corresponding theory. In that case, the theory is just classical and it is not satisfied by $DSR$, $VSR$ and $dSSR$. So the $NC$  may not  be regarded as the deviation of Finsler space-time. This Lagrangian approaches from $NC$ and Finsler geometry can be applied to any black hole, since we could deform any black hole background arrived some modified dispersion relations, one can obtain the equivalency between $NC$ geometry and Finsler geometry.\\
\section{Conclusion}
In this paper, we shortly explained the massive gravity black hole and its Lagrangian. Then, we focused on the $NC$ picture and obtained the corresponding Lagrangian and calculated some canonical relations and momentum  quantities. These quantities helped us to make the Hamiltonian for the mentioned black hole system. In order to apply the $NC$ geometry to the pointed Hamiltonian, we modified such quantities in terms of new variables. The new variables led us to have new Hamiltonian in the form of harmonic oscillator. Then, we applied the $NC$ geometry to the corresponding deformed Hamiltonian and obtained the Lagrangian of massive gravity black hole. Finally, by comparing Lagrangian of massive gravity black hole in $NC$ and Finsler geometries (Eqs. (40) and (45)), we found that they can be the same if $\eta=\frac{(\theta-\beta)}{1-\frac{\theta^{2}}{4}}$ for $\theta^{2}\simeq0$. It means that this similarity guides us to consider both geometries to be equivalent under some considerations. Another proof for this result can be found in the physical concept of the mentioned conditions. As we know, Finsler geometry is related to the momentum component in addition to position (Eq. (34)) and it can be traced in the mentioned condition as the dependence of $\eta$ to $NC$ parameters $\theta$ and $\beta$. When we compare the corresponding Lagrangian in massive gravity without charge for $NC$ and Finsler geometries, the obtained effective potential must be zero. In that case, the kinetic energy of particle near the black hole is smaller than its effective potential. For this reasons, it may be interesting to add some charge and rotation to the massive gravity black hole solution and investigate the relation between two geometries. Also here one can check the mentioned approaches to different black holes with and without charge. It will be also interesting to see the effect of electromagnetic fields on the above equivalency.

\end{document}